\documentstyle[amssymb,preprint,prb,aps]{revtex}

\begin{document}
\draft
\title{Thermoelectric power and transport properties of pure and Al-doped MgB$_{2}$}
\author{B. Lorenz, R. L. Meng, Y. Y. Xue, C. W. Chu$^1$}
\address{Texas Center for Superconductivity and Department of Physics, University of\\
Houston, Houston, Texas 77204-5932, USA}
\address{$^1$also at Lawrence Berkeley National Laboratory, 1 Cyclotron Road,\\
Berkeley, California 94720}
\date{\today}
\maketitle

\begin{abstract}
We have measured the thermoelectric power, S, and resistivity, $\rho$, of
pure and Al-doped MgB$_2$. S is positive and increases linearly with
temperature above the superconducting transition temperature, T$_c$.
Deviations from the linear dependence appear at higher temperature, $T>T_0
\approx 160 K$. T$_c$ and T$_0$ both decrease with Al doping whereas the
slope of S(T) in the linear range increases with the Al content. The data
are discussed in terms of doping induced changes of the Fermi surface and
the density of states at the Fermi level.
\end{abstract}






The recent discovery\cite{Akimitsu} of superconductivity in $MgB_{2}$ at
temperatures as high as 40 K has initiated a tremendous amount of
experimental and theoretical activity with the goal to understand the basic
mechanism of superconductivity in this exciting compound. Two competing
models\cite{Kortus,Hirsch} were proposed to account for the superconducting
properties in $MgB_{2}$ and the high $T_{c}$ of 40 K. While both models
attribute the superconductivity to the boron-sublattice conduction bands,
the pairing mechanisms proposed differ significantly. Kortus et al.\cite
{Kortus} suggested a BCS-type mechanism with strong electron-phonon coupling
and high phonon energy of the light boron atoms. This mechanism is supported
by the observation of an isotope effect on $T_{c}$\cite{Budko}, a BCS-like
superconducting gap structure\cite{Rubio}, and a strong negative pressure
coefficient of $T_{c}$.\cite{Lorenz,Saito} Alternatively, Hirsch\cite{Hirsch}
proposed a ''universal'' mechanism where superconductivity in $MgB_{2}$ is
driven by the pairing of dressed holes. In fact, indications for hole type
conduction in the normal phase were found in the positive thermoelectric
power.\cite{Lorenz} The hole character of carriers was confirmed recently by
Hall measurements and the similarities to high-$T_{c}$ superconductors have
been discussed.\cite{Kang}

Magnetic susceptibility measurements on $Mg_{1-x}Al_{x}B_{2}$\ have shown
that electron doping suppresses $T_{c}$ by a few degree for up to 10 \%
aluminum and superconductivity completely disappears due to a structural
instability for $x>0.1$\cite{Slusky}. This negative doping effect on $T_{c}$
could be explained by both pairing mechanisms discussed above. Within the
BCS model the electron doping gives rise to an increase of the Fermi energy, 
$E_{F}$, and, according to the band structure calculations of $MgB_{2}$,\cite
{Kortus} to a decrease of the density of states, $N(E_{F})$. This will
result in a lower $T_{c}$. In the ''universal'' mechanism,\cite{Hirsch} in
analogy to high-$T_{c}$ cuprates, the electron doping will reduce the number
of hole carriers. In the underdoped regime $T_{c}$ is also expected to
decrease. However, for underdoped high-$T_{c}$ compounds it is well known
that the pressure coefficient of $T_{c}$ is positive, contrary to the data
for $MgB_{2}$.\cite{Lorenz,Saito} Thermoelectric power, $S$, and
resistivity, $\rho $, may give some insight into the normal state conduction
process and the electronic structure. We have, therefore, studied the
temperature dependences of $S$ and $\rho $ of $MgB_{2}$ and the
electron-doped solid solution, $Mg_{1-x}Al_{x}B_{2}$ ($x<0.1$). The data
reveal a linear temperature dependence of $S$ from $T_{c}$ up to $%
T_{0}\thickapprox 160\ K$ typical for the diffusion thermopower of metals.
The positive sign and the low absolute value of several $\mu V/K$ are
characteristic for a hole type metallic conductor. However, the saturation
of $S$ close to room temperature shows that a more detailed consideration of
the structure of the Fermi surface is needed to understand the transport
properties. The slope of the linear part of $S(T)$ changes with $Al$ doping
and $T_{0}$ is reduced indicating changes in the Fermi surface due to the
electron doping.

Polycrystalline $Mg_{1-x}Al_{x}B_{2}$ samples were prepared by solid state
reaction method as described earlier\cite{Lorenz,Slusky}. X-ray powder
diffraction spectra of the samples show a minor amount of MgO as an impurity
phase. The samples were dense enough to be connected with indium pads to
thin platinum wires and thermocouples for resistivity and thermopower
measurements, respectively. The resistivity was measured by the standard
four lead method using an ac resistance bridge, LR 700. For thermoelectric
power measurements we used a home made apparatus and a sensitive ac
technique with an accuracy of 0.1 to 0.2 $\mu V/K$. The room temperature
resistivity, $\rho $, increases from about 70 $\mu \Omega cm$ to 120 $\mu
\Omega cm$ upon doping to 10 \% $Al$ (Fig. 1). This increase is compatible
with the reduction of hole carriers. However, the absolute value of $\rho $
may strongly depend on porosity and grain boundary scattering since the
samples exhibit some porosity. The inset in Fig. 1 shows the resistivity
close to $T_{c}$. All three samples (x=0, 0.05, 0.1) show a sharp resistance
drop with a transition width  $<0.5\ K$. The decrease of $T_{c}$ by about 2
K for $x=0.1$ is in good agreement with the recent magnetization
measurements.\cite{Slusky}

The temperature dependence of the thermoelectric power for $x=0$, $0.05$,
and $0.1$ is shown in Fig. 2. The three curves are separated by an offset of
2 $\mu V/K$ in order to better distinguish the data sets. The inset of Fig.
2 enlarges the superconducting transition region. The transitions are sharp (%
\mbox{$<$}%
0.5 K width). The $T_{c}$ decrease with increasing aluminum content, $x$, is
consistent with the resistivity data and susceptibility experiments.\cite
{Slusky} Within the BCS theory the decrease of $T_{c}$ with increasing
electron number may be explained as a density of states (DOS) effect. The
Fermi energy, $E_{F}$, is close to an edge of rapidly decreasing DOS and any
increase of $E_{F}$ due to $Al$-doping will result in a remarkable decrease
of $N(E_{F})$. In the BCS-description $T_{c}$ is given by the McMillan
formula\cite{McMillan}

\begin{equation}
T_{c}\varpropto \omega \exp \left[ \frac{-1.02\left( 1+\lambda \right) }{%
\lambda \left( 1-\mu ^{\ast }\right) -\mu ^{\ast }}\right]   \eqnum{1}
\end{equation}
where $\omega $ is the characteristic phonon frequency, $\mu ^{\ast }$ the
Coulomb repulsion, and $\lambda =N\left( E_{F}\right) \left\langle
I^{2}\right\rangle /M\left\langle \omega ^{2}\right\rangle $ is the electron
phonon coupling constant. $\left\langle I^{2}\right\rangle $ is the averaged
square of the electronic matrix element, $M$ the atomic mass, and $%
\left\langle \omega ^{2}\right\rangle $ the averaged square of the phonon
frequency. $\lambda $ decreases proportional to $N\left( E_{F}\right) $
resulting in a decrease of $T_{c}$. It should be noted that equation (1) was
also used successfully\cite{Loa} to explain the observed decrease of $T_{c}$
with external pressure.\cite{Lorenz,Saito}

The most interesting features of the thermoelectric power are the positive
sign, the overall small value and the linear dependence in the low
temperature range (indicated by dashed lines in Fig. 2). The positive but
small value is typical for hole type metals. This observation is supported
by the results of band structure calculations. Although the Fermi surface of 
$MgB_{2}$ shows a complex structure with sheets of hole type as well as
electron type states,\cite{Kortus} it appears that the hole states dominate
in the low temperature ($T<T_{0}$) electronic transport. Recent Hall
measurements also provide evidence for predominantly hole type conduction in 
$MgB_{2}$\cite{Kang} and extended band structure calculations have shown
that the positive Hall coefficient is the result of a superposition of
positive and negative components in the polycrystalline sample.\cite{Satta}
Considering the linear $S\varpropto T$ dependence below 160 K, one is
tempted to interpret the low temperature part as the diffusion thermopower, $%
S_{d}$, of hole type metals. In fact, according to the Mott formula the
diffusion thermopower is linear in $T$ if higher order corrections are
neglected:\cite{Blatt} 
\begin{equation}
S_{d}=\frac{\pi ^{2}k^{2}T}{3e}\left[ \frac{\partial \ln \sigma \left(
\varepsilon \right) }{\partial \varepsilon }\right] _{E_{F}}  \eqnum{2}
\end{equation}

$k$ is Boltzmann's constant, $e$ is the charge of the carriers, and $\sigma
\left( \varepsilon \right) $ is a conductivity like function for electrons
of energy $\varepsilon $. At low temperatures, in the residual resistance
region, the carrier relaxation time is limited by impurity scattering and
the logarithmic derivative in equ. 2 is simply $E_{F}^{-1}$ leading to the
expression for $S_{d}$\cite{Blatt}

\begin{equation}
S_{d}=\frac{\pi ^{2}k^{2}T}{3eE_{F}}  \eqnum{3}
\end{equation}

Here $E_{F}$ is the Fermi energy calculated from the edge of the conduction
band. However, it is not expected that this simple formula is valid in the
whole temperature range. First of all, in deriving equation 3 a spherical
Fermi surface was assumed and a $T$-independent relaxation time was adapted.
This approximation limits the application of (3) to low temperatures, i.e.
the residual resistance region. Band structure calculations show that the
Fermi surface of $MgB_{2}$ is far from being spherical and, in particular,
also reflects the anisotropy of the layered structure.\cite{Kortus}
Secondly, the deviation from linearity at $T_{0}$ and the saturation of $S$
close to room temperature cannot be explained by (3). This phenomenon seems
to be related to the complexity of the Fermi surface and the existence of
electron type sheets. These minor carriers may add a negative contribution
to the Seebeck coefficient that increases at higher temperature. For the $Al$
doped samples the crossover temperature, $T_{0}$, clearly decreases to about
130 K ($x=0.05$) and 118 K ($x=0.1$). However, for temperatures $T<T_{0}$
the experimental data perfectly follow the linear relation (3) and the
temperature dependence of resistivity is small (indicating that impurity
scattering is dominating). Assuming that electronic transport in this range
is due to hole carriers in the $\sigma $ bands we can use equation 3 to
estimate the Fermi energy for these $\sigma $ holes (for hole carriers this
energy has to be referenced to the top of the $\sigma $ bands). For $MgB_{2}$
the slope of $S(T)$ below 160 K is $0.042\ \mu V/K^{2}$ and, according to
(3), $E_{F\sigma }=0.57\ eV$. This value is in fair agreement with the
difference between the Fermi energy and the top of the $\sigma $ bands of
about $0.9\ eV$ calculated by Suzuki et al.\cite{Suzuki} for $MgB_{2}$. With
increasing doping the slope of $S(T)$ also increases to $0.047\ \mu V/K^{2}$
(x=0.05) and $0.050\ \mu V/K^{2}$ (x=0.1) indicating a decrease of $%
E_{F\sigma }$ by about $16\ \%$ ($x=0.1$). This decrease is in very good
quantitative agreement with the calculated $17\ \%$ for $%
Mg_{0.9}Al_{0.1}B_{2}$.\cite{Suzuki} The results of this paragraph show that
despite the complex structure of the Fermi surface, the transport properties
of pure and $Al$ doped $MgB_{2}$ in the low temperature range are more
similar to a conventional hole type metal.

It is interesting to note that there is obviously no phonon drag
contribution to the thermoelectric power of $Mg_{1-x}Al_{x}B_{2}$ (Fig. 2).
The phonon drag effect is most common for pure metals and results in an
enhancement of $S$ in the low temperature region. The absence of this
contribution in $MgB_{2}$ has yet to be explained. The predominantly linear
temperature dependence of $S(T)$ is similar to the thermopower of disordered
metals where the phonon heat current is suppressed.\cite{Blatt} Disorder
could be introduced by high porosity, defects, impurities, or the dopants ($%
Al$) itself. In spite of the fact that the linearity $S\varpropto T$ is most
pronounced in the pure $MgB_{2}$ where there is no $Al$ on $Mg$ sites it is
unlikely that the disorder induced by dopants may explain the absence of a
phonon drag contribution to $S$.

We have not considered yet the possible anisotropy of the Seebeck
coefficient. The thermoelectric tensor of hexagonal materials has two
independent coefficients corresponding to measurements made parallel ($%
S_{\parallel }$) and perpendicular ($S_{\perp }$) to the hexagonal axis.
Both coefficients can be quite different, as shown for some hexagonal
metals. \cite{Rowe} Data from polycrystalline samples can only be considered
as an average over all possible grain orientations. For example, $%
S_{\parallel }$ and $S_{\perp }$ of $Zn$ are both nonlinear and of very
different values over a large temperature range, 0%
\mbox{$<$}%
T%
\mbox{$<$}%
300 K. Discussions of the Hall coefficients indicate a strong anisotropy in
different crystallographic directions.\cite{Satta} From the shape of the $%
MgB_{2}$ Fermi surface it could be expected that the Seebeck coefficient is
anisotropic. Note that the hole type areas form cylinders (bonding $p_{x,y}$
bands) running along $\Gamma -A-\Gamma $ and a perpendicular tubular network
(bonding $p_{z}$ bands) whereas the electron type sheets form only a tubular
network in the plane perpendicular to $\Gamma -A-\Gamma $ (antibonding $p_{z}
$ bands).\cite{Kortus} This anisotropy should be more obvious at higher
temperature where the scattering is determined by phonons. In the low
temperature impurity scattering range and in the absence of the phonon drag
contribution the Seebeck coefficient is expected to show less anisotropy. In
the lack of $MgB_{2}$ single crystals it appears difficult to measure the
anisotropy of the thermopower. However, c-axis oriented thin films of $%
MgB_{2}$ may be used to extract the in-plane Seebeck coefficient, $S_{\perp }
$.

In conclusion, the resistivity and thermoelectric power of pure and $Al$%
-doped polycrystalline $MgB_{2}$ have been measured. The decrease of $T_{c}$
with $Al$ doping previously deducted from susceptibility measurements was
confirmed. The Seebeck coefficient of undoped $MgB_{2}$ was found to
increase linearly with temperature from $T_{c}$ to about 160 K. This
increase, the positive sign, and the overall small value of $S$ are
compatible with the assumption that $Mg_{1-x}Al_{x}B_{2}$ is a hole type
normal metal. This result is further supported by the increase of the slope
of $S(T)$ with electron doping. Deviations from linearity at higher
temperatures are discussed in terms of a contribution from electron like
sheets of the Fermi surface. The origin of the missing phonon drag
contribution is still an open question. The Seebeck coefficient of $MgB_{2}$
may be anisotropic and measurements of single crystals or oriented thin
films are required. The current data yield indirect support of the BCS
mechanism for the superconducting transition.


\acknowledgments
This work was supported in part by NSF Grant No. DMR-9804325, MRSEC/NSF
Grant No. DMR-9632667, the T. L. L. Temple Foundation, the John and Rebecca
Moores Endowment, and the State of Texas through the Texas Center for
Superconductivity at the University of Houston; and at Lawrence Berkeley
Laboratory by the Director, Office of Energy Research, Office of Basic
Sciences, Division of Material Sciences of the U. S. Department of Energy
under Contract No. DE-AC0376SF00098.


%
%
\begin{figure}[tbp]
\caption{Resistivity of Mg$_{1-x}$Al$_x$B$_2$. The inset shows the details
near the superconducting transition.}
\label{Fig. 1}
\end{figure}

\begin{figure}[tbp]
\caption{Thermoelectric Power of Mg$_{1-x}$Al$_x$B$_2$. For clarity, the
curves are offset by a constant and the zero values are indicated by a short
line. The inset shows the region of the superconducting transitions.}
\label{Fig. 2}
\end{figure}

%
%

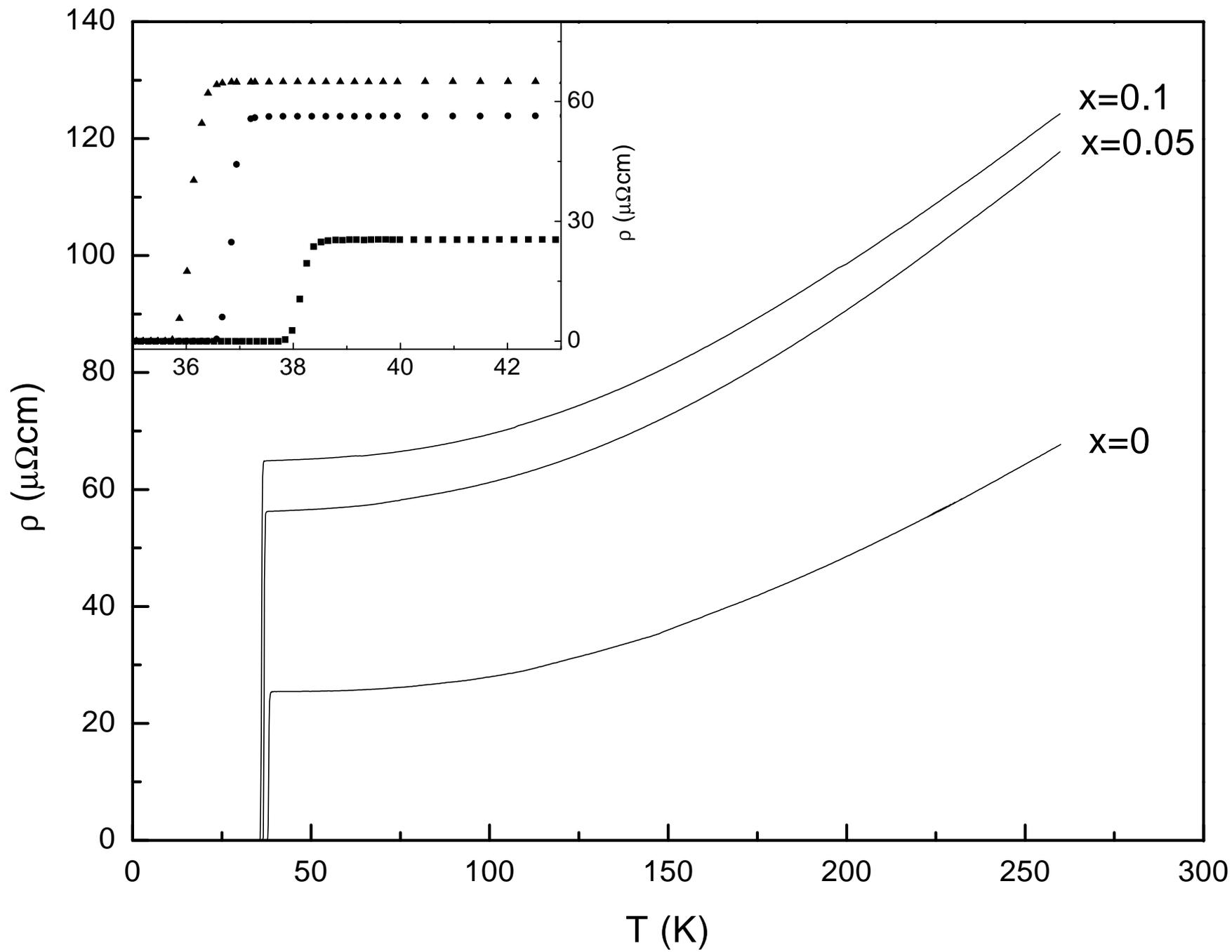
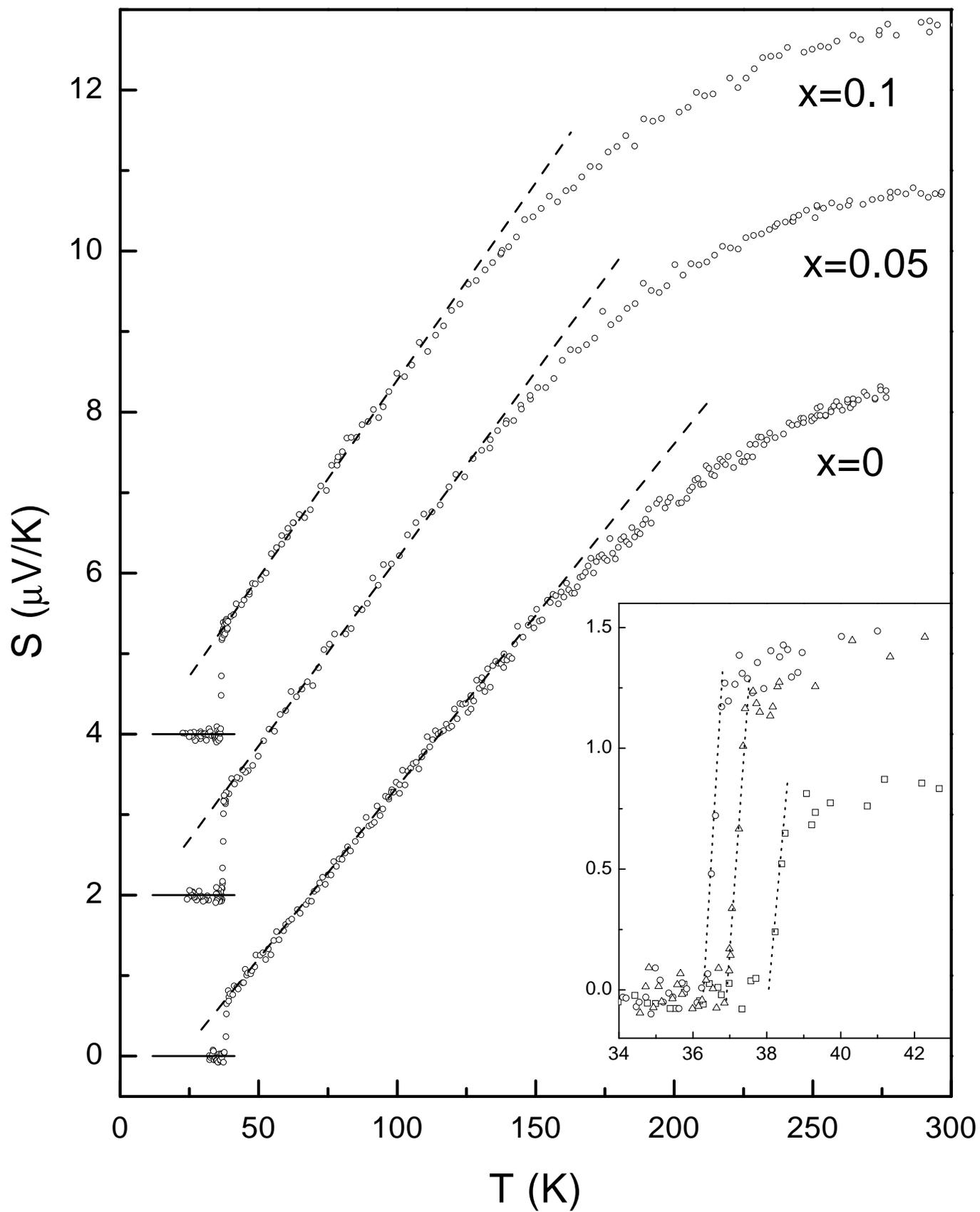
\end{document}